\documentclass[10pt, journal]{IEEEtran}
\IEEEoverridecommandlockouts

\newcommand{\UC}{UBC}

\usepackage{tabularray}

\usepackage{color}
\usepackage{xcolor}

\usepackage{ragged2e}

\usepackage{bbding}
\usepackage{pifont}

\definecolor{bronze}{rgb}{0.8, 0.5, 0.2}

\usepackage{tikz}

\usepackage{amsmath}
\usepackage{amsfonts}
\usepackage{amssymb}
\usepackage{mathtools}
\DeclarePairedDelimiter{\ceil}{\lceil}{\rceil}

\usepackage{xcolor}
\usepackage{circledsteps}
\newcommand\myCircled[2][]{\ifmmode
\Circled[fill color=black,inner color=white,#1]{\mathsf{#2}}
\else
\Circled[fill color=black,inner color=white,#1]{\sffamily#2}
\fi
}

  \renewenvironment{thebibliography}[1]{%
    \begin{oldthebibliography}{#1}%
      \setlength{\parskip}{-0.0ex}%
      \setlength{\itemsep}{-0.0ex}%
  }%
  {%
    \end{oldthebibliography}%
  }

\usepackage{cite}
\usepackage{amsmath,amssymb,amsfonts}
\usepackage{algorithmic}
\usepackage{graphicx}
\usepackage{textcomp}
\usepackage{xcolor}

\def\BibTeX{{\rm B\kern-.05em{\sc i\kern-.025em b}\kern-.08em
    T\kern-.1667em\lower.7ex\hbox{E}\kern-.125emX}}
 
\newcommand{\ignore}[1]{ }

\newcommand*\circled[1]{\tikz[baseline=(char.base)]{
            \node[shape=circle,draw,inner sep=1.0pt] (char) {#1};}}

\usepackage{multirow}
\usepackage{vcell}

\usepackage{comment}

\usepackage{amsmath}
\usepackage{mathtools}

\usepackage{multicol}

\usepackage{enumitem}

\usepackage[hidelinks,colorlinks=true,linkcolor=blue,citecolor=blue]{hyperref}

\usepackage[normalem]{ulem}

\usepackage{makecell}

\usepackage{algorithm}
\usepackage{algorithmic}

\usepackage{colortbl}
\usepackage{hhline}

\usepackage{nccmath}

\usepackage[prependcaption, textsize=footnotesize]{todonotes}

\usepackage{courier}

\usepackage{balance}

\definecolor{lime}{HTML}{A6CE39}
\DeclareRobustCommand{\orcidicon}{%
	\begin{tikzpicture}
	\draw[lime, fill=lime] (0,0) 
	circle [radius=0.16] 
	node[white] {{\fontfamily{qag}\selectfont \tiny ID}};
	\draw[white, fill=white] (-0.0625,0.095) 
	circle [radius=0.007];
	\end{tikzpicture}
	\hspace{-2mm}
}

\foreach \x in {A, ..., Z}{%
	\expandafter\xdef\csname orcid\x\endcsname{\noexpand\href{https://orcid.org/\csname orcidauthor\x\endcsname}{\noexpand\orcidicon}}
}

\newcommand*\titleheader[1]{\gdef\@titleheader{#1}}

\begin{document}

\title{

uHD: Unary Processing for Lightweight and Dynamic   Hyperdimensional Computing
}




\author{
Sercan~Aygun\orcidB{},
Mehran~Shoushtari~Moghadam\orcidA{},
M.~Hassan~Najafi\orcidD{}\\
{School of Computing and Informatics, University of Louisiana at Lafayette, LA, USA}\\       
{\{sercan.aygun, m.moghadam, najafi\}@louisiana.edu}\\

\vspace{-20pt}


}

\markboth{\small{\ \ \ \ \ \ \ \ \ \ \ \ \ \ \ \ \ \ \ \ \ \ \ \   T\lowercase{his} \lowercase{work is accepted to the} D\lowercase{esign,} A\lowercase{utomation and} T\lowercase{est in} E\lowercase{urope} (DATE) C\lowercase{onference} 2024}}%
{\MakeLowercase{\textit{}}}

\titleheader{2016 IEEE 24th International Requirements Engineering Conference}

\maketitle

\begin{abstract}

Hyperdimensional computing (HDC) is a novel computational paradigm that operates on 
long-dimensional vectors known as hypervectors. The hypervectors are constructed as long bit-streams and form the basic building blocks of HDC systems. In HDC, hypervectors are generated from scalar values 
without taking their bit significance into consideration. HDC has been shown to be efficient and robust in various data processing applications, including computer vision tasks. To construct HDC models for vision applications, the current state-of-the-art practice 
utilizes two parameters for data encoding: pixel intensity and pixel position. However, the intensity and position information embedded in high-dimensional vectors are generally not generated dynamically in the 
HDC models. Consequently, the optimal design of hypervectors with high model accuracy requires powerful computing platforms for training. A more efficient approach to generating hypervectors is to create them \textit{dynamically} during the training phase, which results in 
accurate, low-cost, and highly performable vectors. To this aim, 
we use \textit{low-discrepancy} sequences to generate intensity hypervectors only, 
while avoiding position hypervectors. By doing so, the multiplication step in vector encoding is eliminated, 
resulting in a power-efficient HDC system. 
{For the first time in the literature,} our proposed approach 
employs lightweight vector generators utilizing \textit{unary bit-streams} for efficient encoding of data 
instead of using conventional comparator-based generators. 

\ignore{
Our synthesis results show that the proposed architecture provides up to $488\times$ savings in energy consumption and $7\times$ lower hardware area footprint compared to a state-of-the-art HDC design.
}

\end{abstract}


\section{Introduction}
Traditional computing systems based on positional binary radix 
encounter practical limitations in the efficient hardware design of today's big data applications. 
Particularly for cognitive tasks 
with iterative and complex learning procedures, these systems suffer from extremely high power and memory consumption. 
Emerging 
computing technologies 
such as Hyperdimensional Computing (HDC), Stochastic Computing (SC), Unary Bit-stream Computing (UBC), Quantum Computing (QC), and Approximate Computing (AC)  are shaping 
the next generation of computing systems. Among these, HDC has recently gained significant attention due to its lightweight, robust, and efficient solutions 
for various learning and cognitive tasks~\cite{aygun2023learning, 9107175}, particularly for natural language processing~\cite{RahimiGitHubBasedPaper} and image classification~\cite{8801933}. 
HDC encodes information using holographic hyperdimensional vectors, known as \textit{hypervectors},
consisting of randomly distributed binary values of $-1$ (logic-0) and $+1$ (logic-1).
This unconventional representation enables fast, robust, efficient, 
and fully parallel processing of large sets of data~\cite{6807729}.

For high-quality HDC, hypervectors are expected to be \textit{orthogonal}, i.e., uncorrelated with each other. 
By generating 
\textit{pseudo-random} vectors, prior works encode data to hypervectors that are only \textit{nearly} orthogonal. 
This work introduces a novel hypervector encoding scheme, which is radically 
different 
from the encoding methods currently used in HDC systems. 
We propose a simpler and more effective method to achieve orthogonality 
by drawing an analogy between HDC and 
SC~\cite{Alaghi_Survey_2018}. Instead of 
relying on pseudo-randomness, we leverage \textit{quasi-randomness} provided with 
low-discrepancy (LD) 
sequences~\cite{7927069} for generating high-quality hypervectors. In addition, for the first time to the best of our knowledge, we take advantage of 
\UC~and its unary data representation~\cite{9139000} for lightweight design of  HDC systems. 
In what follows, we summarize the primary contributions of this work.
\\
\textbf{\ding{172}} Utilizing
quantized LD 
sequences for hypervector encoding 
for the first time in the  literature. \\
\ding{173} Eliminating position hypervectors in HDC system, 
alleviating the total memory consumption, vector generation load, and arithmetic operations. \\
\ding{174} Developing \textbf{\texttt{uHD}}, a hybrid HDC system 
integrating unary bit-streams and hypervector processing.
\\
\ding{175} Developing a lightweight combinational logic 
to compare unary bit-streams for 
dynamic generation of 
hypervectors.\\ 
\ding{176} A new circuitry for the 
binarization operation needed in HDC systems. \\
\ding{177} Achieving a higher image classification accuracy compared to the baseline HDC with pseudo-random hypervectors. 

\ignore{
This paper is organized as follows. Section~\ref{background_and_motivation} presents the background of HDC systems and the motivation behind this work. Section~\ref{efficient_hypervector_design} presents our new unary HDC system and 
describes the systematic steps of designing a complete HDC hardware system with limited resources. 
Section~\ref{design_evaluation} presents the evaluation results. Finally, Section~\ref{conclusions} concludes the paper.
}

\section{Background and Motivation}
\label{background_and_motivation}
HDC maps 
raw input data into a high-dimensional space with 
hypervectors of $+1$s and $-1$s~\cite{QuantHD}. Each dimension in this space corresponds to a feature or attribute in the original 
data. 
HDC consists of two primary steps: hypervector \textit{generation} and \textit{encoding}, of which the latter creates 
another hypervector. 
While the encoding step has been extensively discussed in 
the literature~\cite{9107175, KleykoPART1}, vector generation is typically left to the performance of pseudo-randomness \cite{HV_Design}. 
When a \textbf{scalar} value $X$ is to be represented using 
a hypervector, its numerical value can be used 
for vector generation. However, when $X$ is a \textbf{symbolic} data, 
(e.g., a \textit{letter}) a \textit{proper} vector should be attributed to the symbol. The term 
\textit{proper} emphasizes the importance of orthogonality, as each symbol without numerical information should be treated equally and embedded in hypervectors without any bias towards one symbol over another. In other words, each hypervector should have an equal number of $+1$s and $-1$s 
with an independent random distribution. This representation requires a good randomness 
to ensure 
hypervectors remain uncorrelated with each other. 
An important target of this work is to produce hypervectors with ideal 
orthogonality.  
For the scalar case, 
$X$ can be a grayscale pixel value \cite{8801933} ( 
{\footnotesize $0 \leq X \leq 255$} for 8-bit representation), the amplitude of a discrete signal \cite{9107175}, or a numerical feature of data \cite{QuantHD}. This work will follow the convention for image classification, so we assume {\footnotesize $X$} is a pixel value.

\begin{figure}[t]
  \centering
  \includegraphics[width=\linewidth]{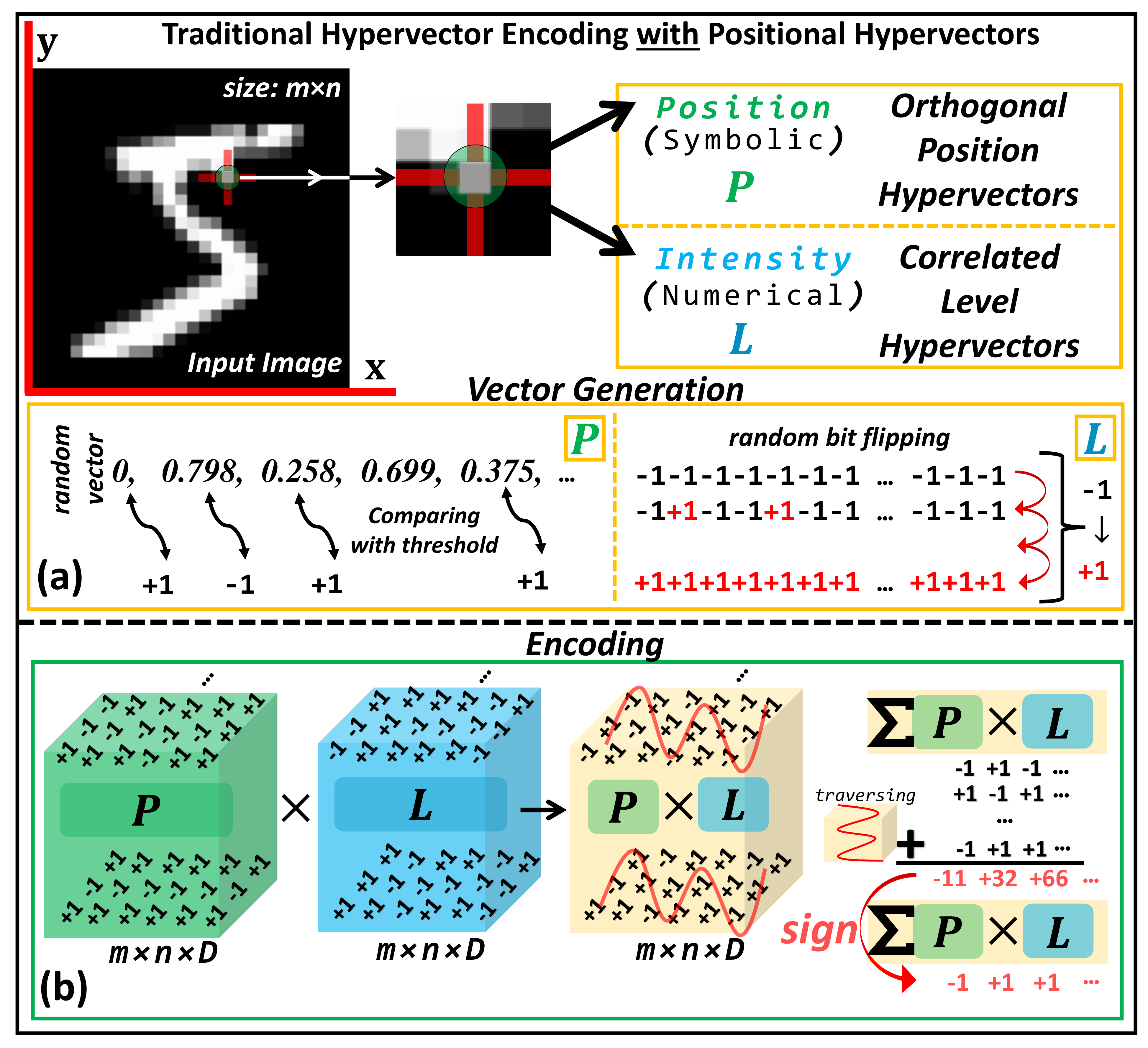}
\vspace{-1.5em}  \caption{Traditional hypervector design and encoding. (a) \textit{Positional} and \textit{Level} hypervector generation, (b) Binding, Bundling, and Binarization. 
}
\vspace{-.5em}
  \label{conventinal_steps}
\end{figure}

Fig.~\ref{conventinal_steps}(a) shows a sample image pixel and 
its position ($\boldsymbol{P}$) and  level ($\boldsymbol{L}$) hypervectors.
Hypervectors are assigned with a dimension or size of $D$. $\boldsymbol{P}s$ are obtained from \textbf{symbolic} data, 
and $\boldsymbol{L}s$ 
from \textbf{scalar} values. $\boldsymbol{P}s$ are generated by comparing 
random ($R$) numbers ({\footnotesize $0 \leq R_{1..D} \leq 1$}) 
and a threshold value ($t=0.5$; \textit{no-bias} point between $0$ and $1$). 
$\boldsymbol{L}s$ are typically generated by \textit{bit flipping} \cite{HV_Design}. 
Similar to comparing $t=0.5$ and random numbers, {\footnotesize $0 \leq R \leq D$}, each {\footnotesize $R$} and threshold {\footnotesize $t = k \times \frac{D}{2^n}$} are compared to generate one dimension of $\boldsymbol{L}$ 
at the {\footnotesize $k^{th}$} cycle, where {\footnotesize $D \geq 2^n$} ($n$-bit precision, {\footnotesize $k_{max}$=$2^n$}). Hence, closer numerical scalars have similar hypervectors, while different numerical scalars have 
more uncorrelated hypervectors. In generating both $\boldsymbol{P}$ and $\boldsymbol{L}$, 
the comparison returns a $+1$ or $-1$ 
value for 
any hypervector position. 
If {\footnotesize $R > t$}, the corresponding 
position is set to 
$-1$; otherwise, it is set to $+1$ \cite{9107175}.

Fig.~\ref{conventinal_steps}(b) illustrates the remaining 
encoding steps on 
the generated hypervectors. This step composes 
the \texttt{class hypervector} ($\boldsymbol{C}$), which holds the overall representation of a class (e.g., Fig.~\ref{conventinal_steps} shows an image from class-\textit{5} and its contribution to the corresponding \texttt{class hypervector}). 
Each image in the training set contributes to building the \texttt{class hypervectors} by processing the hypervectors ($\boldsymbol{P}$ and $\boldsymbol{L}$) of all its pixels. 
The generated hypervectors are first multiplied element-wise (via bit-wise \texttt{XOR}). 
This is known as 
\textit{binding}. 
To accumulate the multiplied hypervectors coming from each pixel ($\boldsymbol{L} \oplus \boldsymbol{P}$), positions are traversed. Hypervectors are added to each other 
by another element-wise processing (bit-wise \texttt{popcount}). 
This is known as 
\textit{bundling}~\cite{8490896}. Then, the final values are evaluated for \texttt{class hypervectors} after scanning all data samples of the same class. Finally, the \textit{binarization} operation is performed via a \texttt{sign} function (thresholding with a \texttt{comparator} or a \texttt{subtractor})~\cite{RahimiGitHubBasedPaper}. 
For each class in training, the labeled data are processed to build 
its corresponding \texttt{class hypervector}. 
This operation is performed only once, different from the conventional learning systems having iterative forward passes throughout the batches and epochs.

When all \texttt{class hypervectors} are defined ($\boldsymbol{C}_{1..q}$ with $q$-class dataset), 
the inference step 
measures the accuracy of the testing dataset. 
The same encoding steps
are followed for any 
testing data to obtain a testing hypervector ($\boldsymbol{C}_{test}$). The final classification is performed using a similarity check 
between 
$\boldsymbol{C}_{test}$ vs. $\boldsymbol{C}_{1}$, $\boldsymbol{C}_{2}$, ..., and $\boldsymbol{C}_{q}$. 
In this work, 
we use cosine similarity. 
The highest similarity 
between $\boldsymbol{C}_{test}$ and one of the trained classes gives the classification decision \cite{RahimiGitHubBasedPaper}.

Generating 
pseudo-random hypervectors with high orthogonality during training can be very time- and memory-consuming. 
To obtain a high 
classification accuracy, the best performing $\boldsymbol{P}$ and $\boldsymbol{L}$ random hypervectors are assigned iteratively. 
Hypervectors with different distributions are generated iteratively to find those with the highest orthogonality. 
One of our goals in 
this work is to minimize the number of vector operations. 
The bit-wise \texttt{XOR} operations in the binding process involve 
both $\boldsymbol{P}$ and $\boldsymbol{L}$ hypervectors. 
{We use an encoding for level hypervectors that does not need iteration and provides accurate encoding deterministically \cite{10137195}.} Instead of pseudo-randomness, we provide high orthogonality via \textit{quasi-randomness}~\textbf{\ding{172}}. 
Our approach eliminates the need for position encoding and their corresponding multiplications~\ding{173}. 
Thus, single-iteration vector optimization is guaranteed thanks to the properties of 
LD sequences~\cite{Sobol_TVLSI_2018}.

As the first work of its kind, 
we 
use \textit{unary bit-streams} in HDC systems~\ding{174}.  
UBC utilizes unary (
aka thermometer) coding to represent data using bit-streams with 
logic-$1$s (or logic-$0$s) aligned to the beginning or end of the bit-stream. 
For instance, {\footnotesize $\begin{smallmatrix} X1 & \rightarrow & \textcolor{orange}0 & \textcolor{orange}0 & \textcolor{orange}0 & \textcolor{orange}0 & \textcolor{orange}0 & \textcolor{blue}1 & \textcolor{blue}1 \\ X2 & \rightarrow & \textcolor{orange}0 & \textcolor{orange}0 & \textcolor{blue}1 & \textcolor{blue}1 & \textcolor{blue}1 & \textcolor{blue}1 & \textcolor{blue}1  \end{smallmatrix}$} are two unary bit-streams of size $N$=$7$ representing $2$ and $5$. 
UBC can be exploited for lightweight design of HDC systems. Hypervector generation in current HDC systems requires 
conventional binary \texttt{comparators}, 
which is complex 
and consumes significant power. 
We employ UBC to design 
a new 
lightweight \texttt{comparator} logic 
for dynamic hypervector generation 
\ding{175}.

After optimizing vector generation and minimizing the operations in encoding, we also improve the hardware design for the final stage with accumulation and binarization.  
We propose a concurrent binarization during \texttt{popcount}ing; 
Processing over binary data allows using \texttt{popcount} to only count the number of 
logic-$1$s. The binary output is obtained after $D$ cycles 
to be compared or subtracted from a \textit{threshold} value. 
This requires a separate module for thresholding or subtraction. 
We simplify the binarization module to make the decision on the spot while performing popcount~\ding{176}. 

\ignore{
\section{uHD}
\label{unaryHD_section}
Fig.~\ref{design_consideration} compares important design features of \textbf{\texttt{uHD}} and the baseline HDC system in terms of \textit{hardware complexity}, \textit{dynamicity}, and \textit{offline training independency}.
\textbf{\texttt{uHD}} especially distinguishes itself from the prior art by 
dynamic hypervector generation during training. 
\begin{figure}[t]
  \centering
  \includegraphics[width=\linewidth]{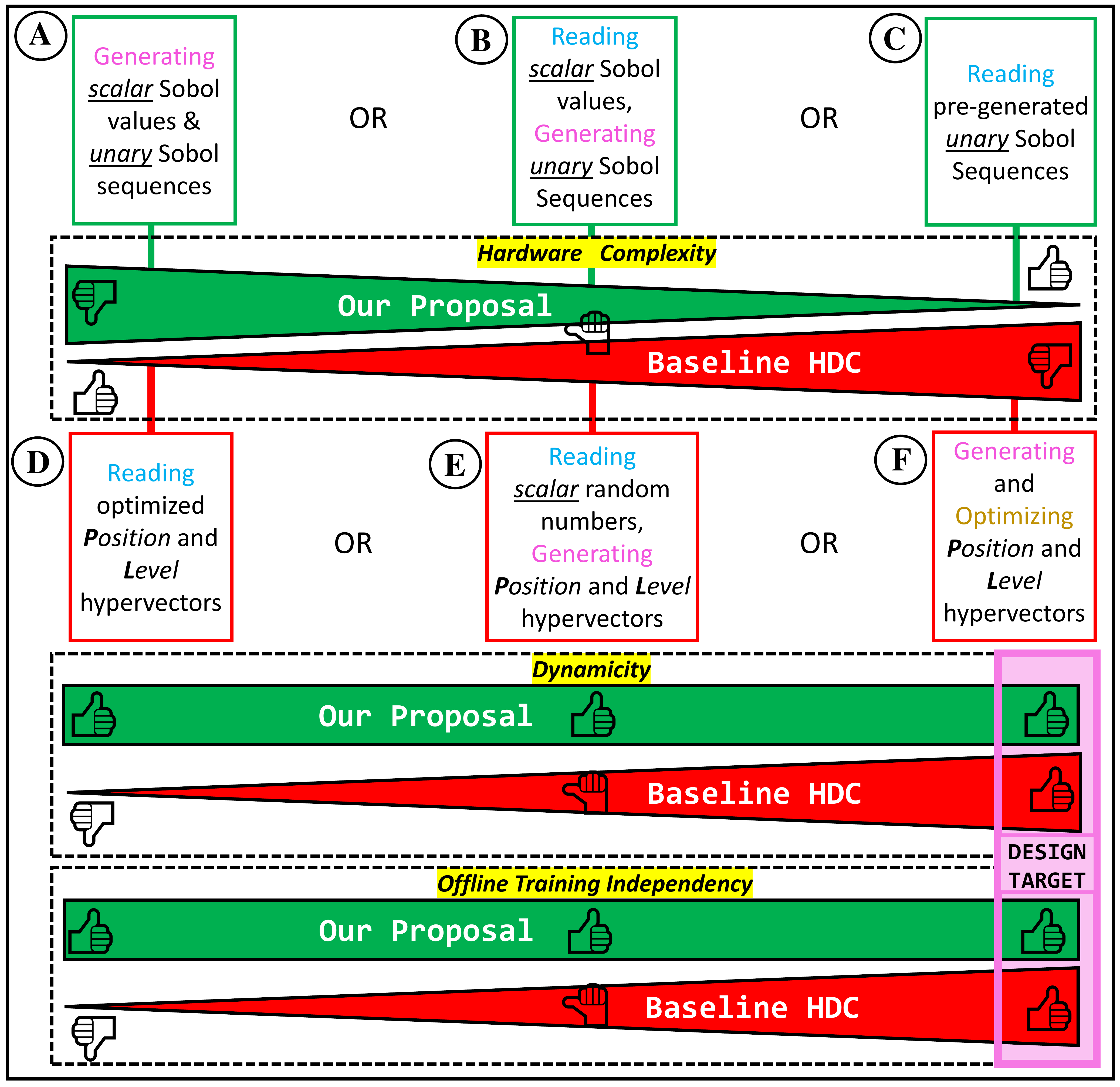}
  \vspace{-2em}
  \caption{Design considerations for \textbf{\texttt{uHD}} and Baseline HDC.}
  \label{design_consideration}
  \vspace{-1em}
\end{figure} 
Our proposed HDC system
uses LD \textit{Sobol} sequences and unary bit-streams 
for vector generation, while baseline HDC requires pseudo-random numbers 
to generate 
hypervectors. 
Due to the large size of hypervectors, generating random values in hardware is costly. 
\textbf{\circled{\scriptsize A}} Generating both scalar and unary Sobol sequences in hardware is complex and so very resource-consuming. 
\textbf{\circled{\scriptsize B}} One solution is to read pre-generated sequences from memory.
\textbf{\circled{\scriptsize C}} Reading pre-generated Sobol sequences in \textbf{unary} format instead of saving Sobol scalars and then converting Sobol scalars to corresponding unary bit-streams gives a lower hardware complexity. 
However, unary bit-streams are longer 
than the conventional binary representation (e.g., a 4-bit binary number is represented with $N$=16-bit unary bit-stream), and memory can be a bottleneck in this case. \textbf{\texttt{uHD}} takes
a different approach. 
It reads Sobol scalars from a random access memory (\texttt{RAM}) and then generates unary sequences by another 
reading from an associative memory  (\textit{fetching}). 

Fig.~\ref{design_consideration} also highlights that the proposed approach is dynamic and independent of offline training and hardware complexity. 
In particular, hypervector generation in \textbf{\texttt{uHD}} is single-iteration and deterministic. 
Even more, \textbf{\texttt{uHD}} is universal; The same design with the same hypervectors can be applied to different-scale (data size, feature size, symbol counts, etc.) applications using the same Sobol scalars. 
Unlike, in baseline HDC design, \textit{dynamicity} and \textit{offline training} vary depending 
on the hardware complexity 
for the needed randomness for $\boldsymbol{P}$ and $\boldsymbol{L}$ generation. \textbf{\circled{\scriptsize D}} Direct hypervector reading from memory is relatively lightweight as it does not need additional hardware for hypervector generation. 
However, already-optimized hypervector usage is a bottleneck for memory due to the dimension 
size. More importantly, a complete architecture for training on edge is not independent and is based on previous optimizations. \textbf{\circled{\scriptsize E}} Reading scalars and generating hypervectors using hardware is a relaxed solution for dynamicity; however, hardware 
complexity starts to be an issue when optimizing 
hypervectors in the hypervector generation phase. Finally, \textbf{\circled{\scriptsize F}} generating scalars and optimizing hypervectors on edge is a fair approach for dynamic and independent design in baseline HDC. For the same dynamicity and independency level, \textbf{\texttt{uHD}}'s memory-based solution 
has a 
mild-level hardware complexity. 
It reads Sobol scalars and generates (fetches) unary bit-streams on-the-fly. 

}

\section{Efficient Hypervector Encoding with \textbf{\texttt{uHD}}}
\label{efficient_hypervector_design}

We call the new unary HDC system \textbf{\texttt{uHD}}. \textbf{\texttt{uHD}} enjoys a lightweight architecture 
by taking advantage of unary processing. It also provides a higher accuracy by exploiting the uncorrelation and recurrence properties of LD sequences. 

\begin{figure}[t]
  \centering
  \includegraphics[width=\linewidth]{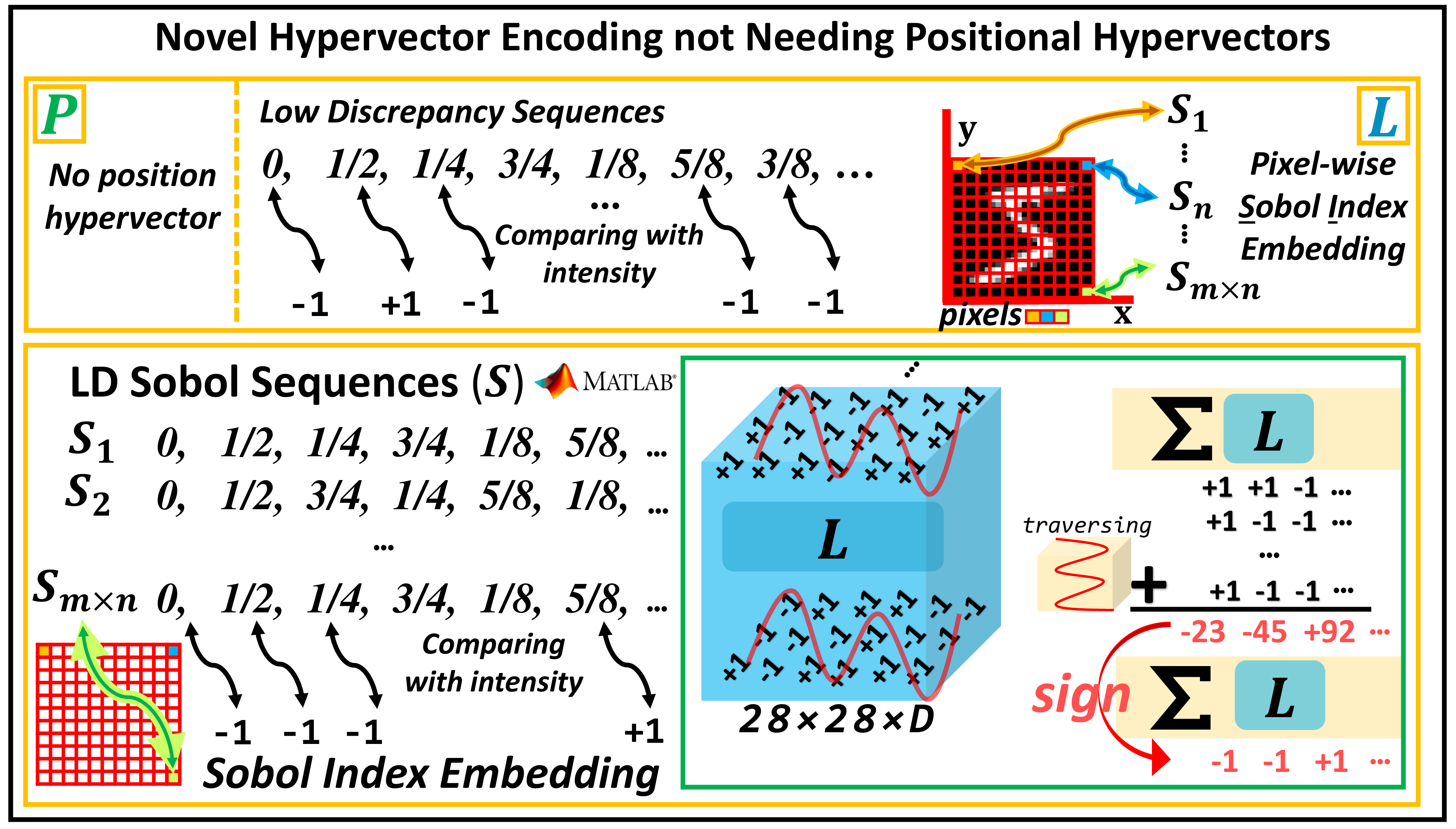}
  \vspace{-1.5em}
  \caption{Hypervector 
  generation using Sobol sequences~\cite{ESL_of_ours}.
}
  \label{propose_sobol_index}
\end{figure}

\textbf{\texttt{uHD}} radically alters the encoding approach 
in HDC systems. Conventional HDC systems are bounded by the spatial information of discrete data. 
LD sequences provide built-in indexes to be used for the positional information. Fig.~\ref{propose_sobol_index} depicts the encoding using LD Sobol~\cite{7927069} scalars and indexes. We eliminate 
$\boldsymbol{P}s$ and only encode $\boldsymbol{L}s$ 
by using Sobol scalars. 
As shown in Fig.~\ref{propose_sobol_index}, for image data encoding, we compare
LD Sobol sequences ($\boldsymbol{S}_i$) (from MATLAB built-in Sobol generator) 
with image intensity values. We do not encode positions; instead, we use the corresponding \underline{i}ndex of any Sobol sequence ($\boldsymbol{S}_i$) ranging from $\boldsymbol{S}_1$ to $\boldsymbol{S}_{\texttt{row} \times \texttt{column}}$. Finally, the non-binary image hypervector formula turns into {\footnotesize $\Sigma_{i=1}^{N} (\boldsymbol{L}_i)$. } Any pixel intensity is encoded based on the pixel position corresponding to the Sobol index. The normalized intensity value (by $D$) is compared with each element in the corresponding Sobol sequence. If the normalized intensity is smaller than the Sobol number, the hypervector position gets $-1$; otherwise, it gets $+1$. After obtaining $\boldsymbol{L}$, we perform the accumulation 
without the encoding's multiplication step. 
Thus, our novel approach achieves a 
\textit{multiplier-less} vector encoding for 
HDC. 

Table~\ref{software_table} compares the performance of the proposed and the baseline HDC with software simulation on an ARM-based embedded processor.  
We run the low-level \texttt{C} language implementation of the two architectures on a resource-constrained ARM1176JZF-S (700 MHz, 32-bit single core, 250 MB RAM).
We compare the two designs in terms of \textit{runtime}, \textit{dynamic memory footprint}, and 
\textit{code size in memory} for processing  each input image. 
The processing runtime of a single image reduces to $0.016$ sec for $D=1K$ (1024), and to $0.058$ sec for $D=8K$, yielding $43.8\times$ and $102.3\times$ speed-up, respectively. In terms of dynamic memory allocation, the new encoding scheme using Sobol LD sequences exhibits $10.4\times$ less memory for $1K$-long 
and $23.6\times$ 
less memory for $D=8K$ 
hypervectors. The memory usage was reduced by $5$ $KB$ with the deployed code of the proposed HDC system.  

From efficient hypervector design to complete hardware modules, we focus on the extended design perspectives for efficient 
HDC system design. Most prior works present hardware design for the inference. However, training on edge devices is a more challenging task. Since for high accuracy, the baseline HDC requires iterative hypervector generation and processing, 
single-pass data processing can significantly reduce the runtime and 
energy consumption. 
Our proposed hypervector encoding achieves this benefit with a 
deterministic and reliable one-time iteration.

\begin{table}
\centering
\caption{Performance 
on an ARM-based Embedded Platform~\cite{ESL_of_ours}}
\renewcommand{\arraystretch}{1.2}

\begin{tabular}{|c|c|c|c|l|c|} 
\cline{1-4}\cline{6-6}
\multicolumn{2}{|c|}{\textbf{\textbf{\textit{Performance - Embed.}}}} & \textbf{Runtime} & \textbf{Dyn. Mem.} & \multicolumn{1}{c|}{} & \multicolumn{1}{l|}{\textbf{Code Mem.}} \\ 
\cline{1-4}\cline{6-6}
\multirow{2}{*}{$1K$} & \textbf{Baseline HDC} & 0.701 sec & 8,496 KB &  & \textbf{\textbf{Baseline~}} \\ 
\cline{2-4}\cline{6-6}
 & \textbf{Our proposal} & 0.016 sec & 816 KB &  & 13.2 KB \\ 
\cline{1-4}\cline{6-6}
\multirow{2}{*}{$8K$} & \textbf{\textbf{Baseline HDC}} & 5.938 sec & 52,401 KB &  & \textbf{\textbf{Ours}} \\ 
\cline{2-4}\cline{6-6}
 & \textbf{\textbf{Our proposal}} & 0.058 sec & 2,220 KB &  & 8.2 KB \\
\cline{1-4}\cline{6-6}
\end{tabular}
\label{software_table}
\end{table}

\begin{figure}
  \centering
  \includegraphics[width=\linewidth]{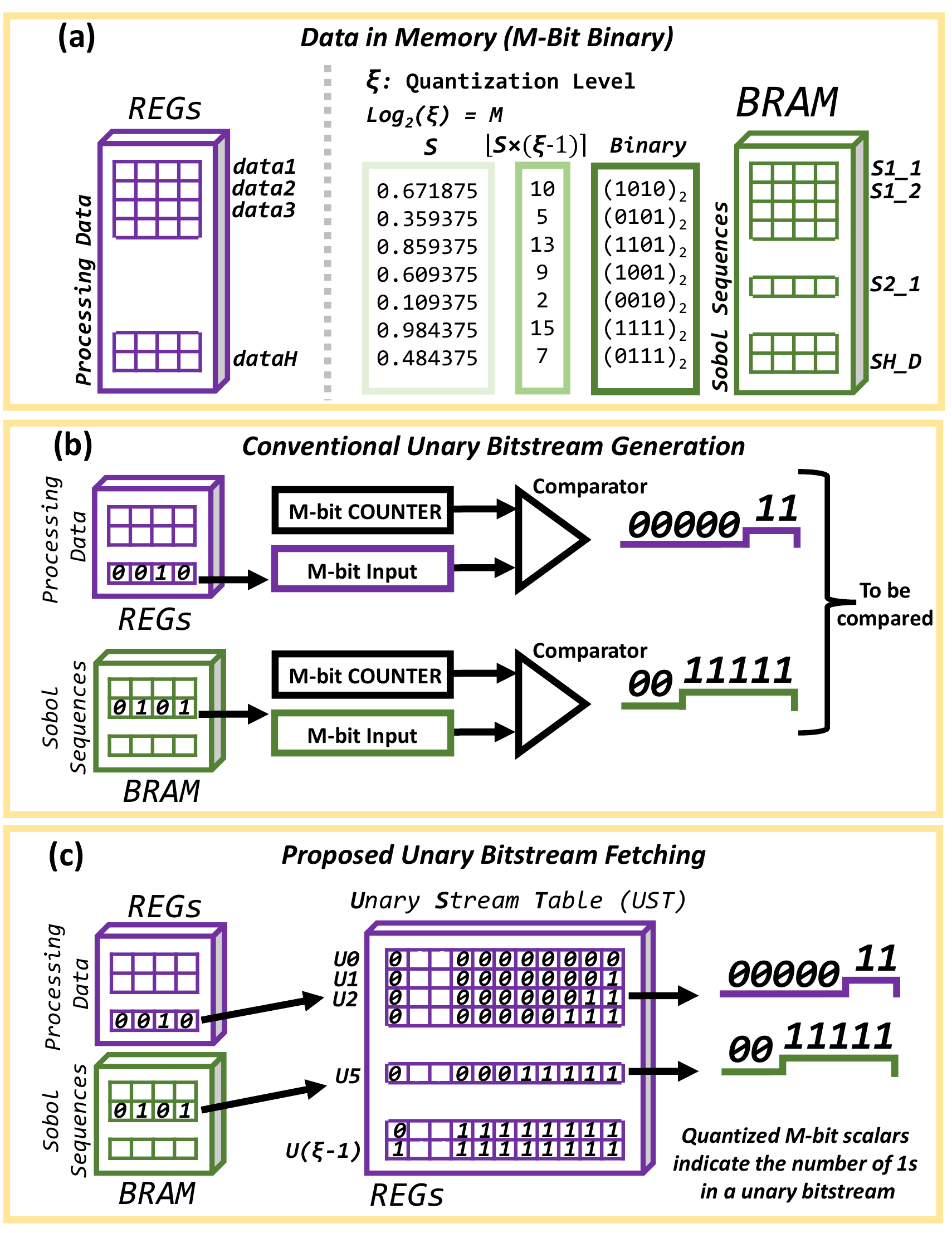}
  \vspace{-1.5em}
  \caption{Getting ready to hypervector generation. (a) Data represented in memory for unary bit-stream processing, (b) Conventional unary stream generation, and (c) proposed associative stream fetching.}
  \label{data_and_sobol_a_b_c}
\end{figure}

\textbf{\texttt{uHD}} reads two sets of data from memory:
(i) \textit{processing data} such as image pixels or features and (ii) \textit{Sobol sequences}. In the proposed approach, we quantize 
both input data and Sobol scalar values. 
Using Sobol scalar and index encoding removes the need for $\boldsymbol{P}s$ and corresponding multiplications. 
We further utilize 
unary bit-streams instead of the conventional binary radix encoding, 
bringing UBC into HDC systems for the first time. Now, let us take a look at the overall 
design. 
With $M$-bit quantization, only $M$-bit data is saved in memory.
The 
input data size depends on 
the \textit{features} or \textit{raw data} size, such as image's \texttt{row}$\times$\texttt{column}. Each Sobol sequence has a length of 
$D$ (i.e., has $D$ Sobol numbers), where $D$ is typically in the range of $1K$ to $10K$. 
Storing all Sobol data in registers 
may not be possible as they may exceed the memory size 
of the resource-constrained 
devices. Therefore, we use block \texttt{RAM} (\texttt{BRAM}) in a re-configurable design platform 
to store 
the quantized Sobol data. 
The processing data are relatively lighter in size, so we keep them in registers. 
Fig.~\ref{data_and_sobol_a_b_c}(a) illustrates how we keep the data. 
The 
data are in quantized binary format (e.g., $M$=4) in registers (\texttt{REG}s). For Sobol scalars, a \texttt{BRAM} module holds binary values in $M$=4-bit (holding $N$=16-bit unary bit-streams. Each scalar shows the total count of logic $1$s in the stream). An example of 
quantizing Sobol sequences is shown in Fig.~\ref{data_and_sobol_a_b_c}(a). Here, $\xi$=16-level quantization is applied to obtain the to-be-saved binary values. 
We note that this data quantization does not affect the accuracy of the system. Even though hypervector generation may experience 
some flipped bits ($+1$ instead of $-1$, or vice versa), 
the accumulated values yield large scalars (non-quantized \texttt{class hypervector}), and the \texttt{sign} of accumulation is not easily affected. 

\begin{figure}
  \centering
  \includegraphics[width=\linewidth]{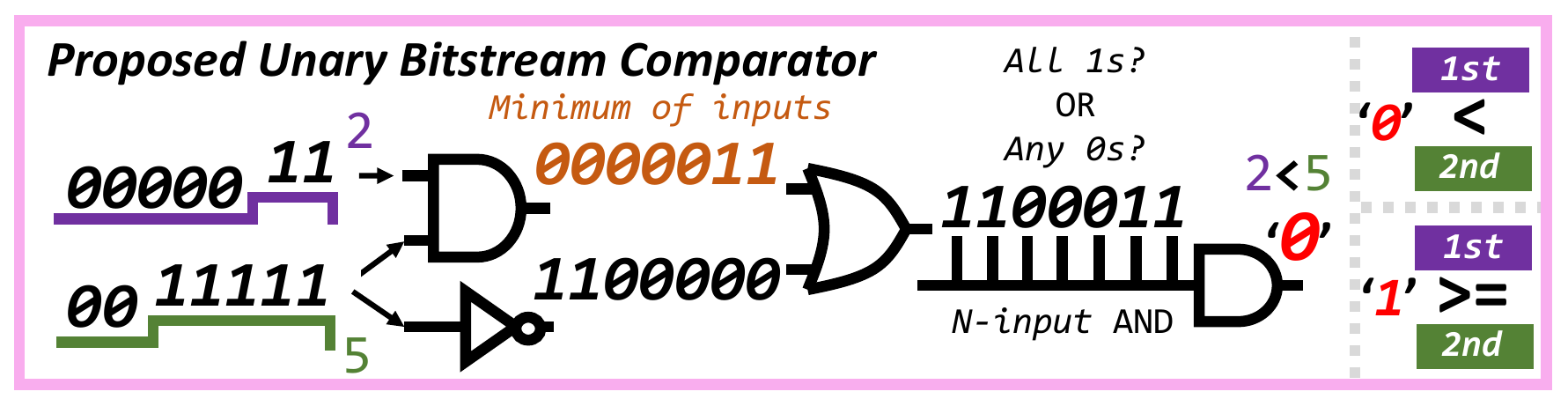}
  \vspace{-1.5em}
  \caption{Proposed comparator of unary bit-streams.}
  \label{proposed_comparator}
\end{figure}

Let us now discuss 
how we convert the data to unary bit-streams. 
Unary bit-streams are conventionally generated by using a pair of $M$-bit  binary \texttt{counter} and \texttt{comparator}~\cite{najafiSortingUnary} as shown in Fig.~\ref{data_and_sobol_a_b_c}(b). 
This design is compact, especially for dynamic bit-stream generation with large sizes. However, our HDC 
design works only on $N$=16-bit 
sequences, and all possible sequences 
can be saved in memory.
The data in memory are converted to unary bit-streams on-the-fly. 
Fig.~\ref{data_and_sobol_a_b_c}(c) shows how we fetch the pre-stored unary bit-streams from an associative memory. 
The binary scalar in \texttt{REG}s or \texttt{BRAM} points to the corresponding 
index of a \underline{U}nary \underline{S}tream \underline{T}able (UST), and the target bit-stream is fetched. We put the first design checkpoint here~\ding{202} 
and compare energy consumption. 
We synthesize the designs using Synopsys Design Compiler with a 45-$nm$ cell library.
We compare the energy consumption of the two approaches for generating one bit of the hypervector. 
\textbf{\texttt{uHD}} consumes $0.77 fJ$ energy while 
the baseline design consumes $0.167 pJ$ (both designed for $D=1K$).

\begin{figure*}
  \centering
  \includegraphics[width=\linewidth]{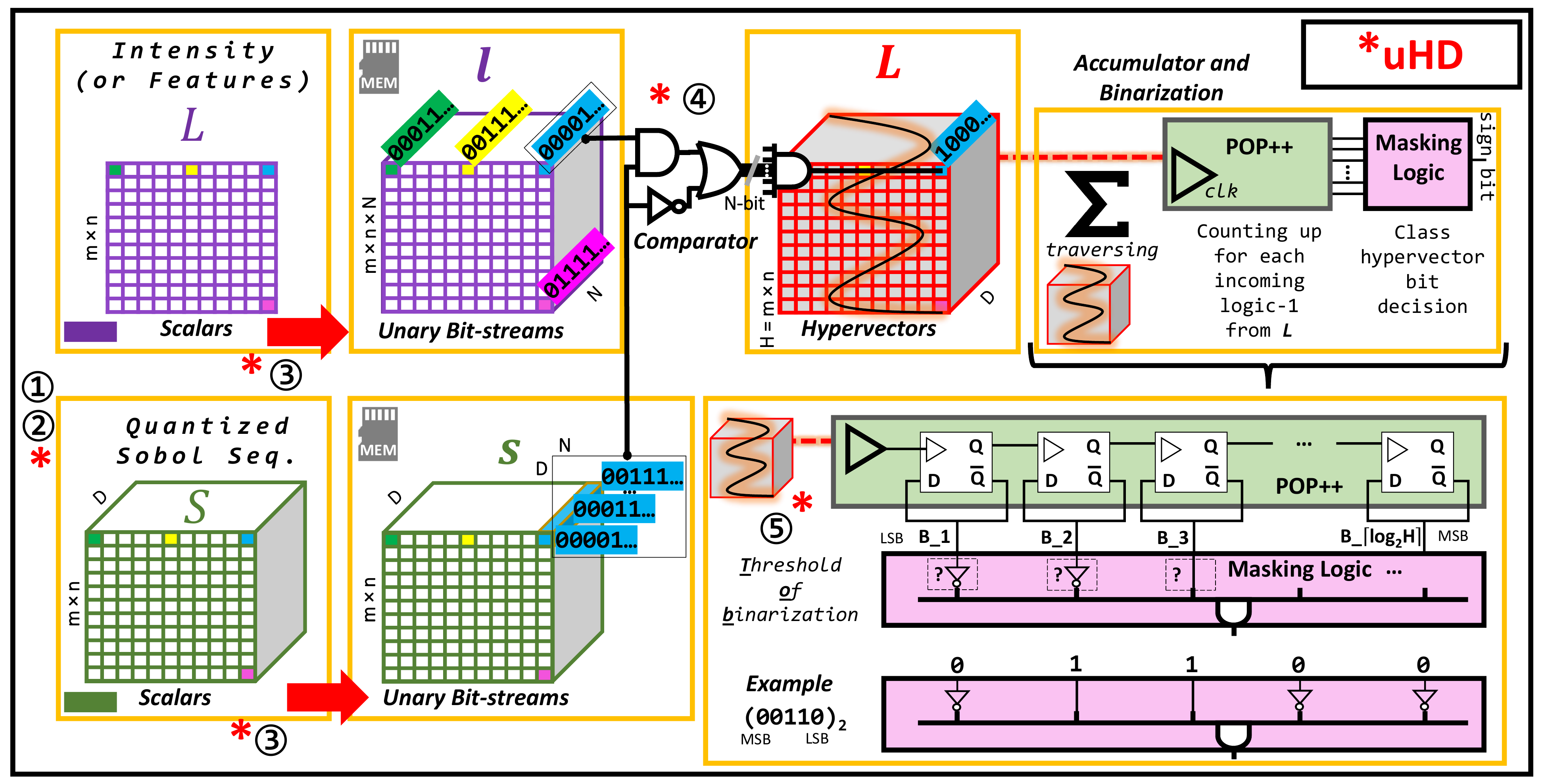}
  \vspace{-1.5em}
  \caption{Complete system overview of \textbf{\texttt{uHD}}. \textcolor{red}{\textbf{\Large{*}}} indicates our proposals.}
  \label{all_of _all}
\end{figure*} 

In our approach, the hypervectors are generated by performing comparison operations between the data and Sobol scalars. 
Instead of directly comparing quantized scalars 
via conventional \texttt{comparator}s, we use a novel unary bit-stream \texttt{comparator}. 
Unary bit-streams (with the same length) are 
correlated, 
and \texttt{AND}ing them 
gives the minimum bit-stream. 
Fig.~\ref{proposed_comparator} illustrates the proposed unary \texttt{comparator} for comparing two $N$=7-bit unary inputs. 
The data- (\includegraphics[width=0.25in]{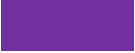}) and Sobol 
(\includegraphics[width=0.25in]{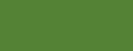}) unary bit-streams are compared to generate one bit of the hypervector: 
If the \texttt{1st} operand  
(here, \textit{data}) is greater than or equal to the  \texttt{2nd} operand (here, \textit{Sobol value}), then the corresponding output is logic-$1$; otherwise, it is logic-$0$. The bit-wise \texttt{AND} operation determines the \textit{minimum} input. 
The inverted unary Sobol is checked (by an \texttt{OR} gate) to see if 
the \textit{minimum} equals this. 
In the example of Fig.~\ref{proposed_comparator}, we compare two unary bit-streams corresponding to values $2$ and $5$.  
The \textcolor{bronze}{\textit{{minimum}}} (bronze color in Fig.~\ref{proposed_comparator}) is decided, and the inverted Sobol is checked if bit-wise \texttt{OR}ing gives all-$1$s or at least one logic-$0$ at the output. If the Sobol number is the \textit{minimum}, the \texttt{OR} operation always returns a logic-$1$, and \texttt{AND}ing $N$ consecutive bits 
at the output guarantees a logic-1 bit for the  hypervector. 
If there is at least one logic-$0$ after \texttt{OR}, the final \texttt{AND} detects it and resets the bit of the hypervector. Since in this example, 
the data (value of $2$ in Fig.~\ref{proposed_comparator}) is less than the Sobol number (value of $5$ in Fig.~\ref{proposed_comparator}), the output is not all-$1$s and a logic-$0$ is generated. 
Here we have the second design checkpoint~\ding{203} 
to evaluate the energy consumption of hypervector generation when using the proposed unary comparator. 
The baseline HDC with conventional comparators consumes $2.49 pJ$, but \textbf{\texttt{uHD}} consumes $0.24 pJ$ (both designed for $D=1K$).

When the level hypervectors ($\boldsymbol{L}s$) are encoded using the steps above, 
the next step is 
the \textit{accumulation} and \textit{binarization}. Fig.~\ref{all_of _all} illustrates the overall 
design of 
\textbf{\texttt{uHD}}. 
The main 
contributions are underscored in this figure. The red-colored $\boldsymbol{L}$ 
is traversed for accumulation. For each hypervector bit, a \texttt{popcount} operation returns a binary output counting the number of logic-$1$s. 
The \texttt{D}-type flip-flops are used in our design for this purpose. 
We propose a new binarization method that operates concurrently with  
the \texttt{popcount} instead of using an extra \texttt{subtractor} or \texttt{comparator} for the \underline{T}hreshold \underline{o}f \underline{B}inarization ($TOB$). The size of the incoming data (the total pixel or feature count), $H$, is the maximum value to count up. Hence, a {\footnotesize $\ceil{log_{2}H}$}-bit counter is required. {\footnotesize $\frac{H}{2} = TOB$} is the critical threshold reached by \texttt{popcount} output for the decision on the logic-$1$s-in-majority. 
When 
the threshold is reached, the \texttt{sign bit} is set for the corresponding bit of the \texttt{class hypervector}; otherwise, it is $0$. 
We propose to use a masking logic for capturing $TOB$, which is in binary: {\footnotesize $(B_{\ceil{log_{2}H}}...B_2 \ B_1)_{2}$}. As shown in Fig.~\ref{all_of _all}, the masking logic is hardwired,  
feeding {\footnotesize $\ceil{log_{2}H}$} bits to the \texttt{AND} gate; 
When \texttt{popcount} reaches $TOB$, this 
hardwired threshold guarantees 
logic-$1$ at the output of \texttt{AND}; otherwise, it remains logic-$0$ \cite{sercanBNN}.  
Here, we set a new design checkpoint~\ding{204} 
and compare the energy consumption of the baseline and \textbf{\texttt{uHD}} design 
for \textit{accumulate-and-binarize} operation. We observe that \textbf{\texttt{uHD}} consumes $34.7 pJ$ energy per feature of the incoming image, while for the same data the baseline design consumes 
$68.7 pJ$ energy 
(both designed for $D=1K$).

\begin{table*}[t]
\centering
\caption{Energy Consumption and $Area \times Delay$ comparison of 
\textbf{\texttt{uHD}} and the Baseline HDC for each HyperVector ({HV}) and image }

\renewcommand{\arraystretch}{1.2}

\begin{tabular}{|c|c|c|c|c|c|c|} 
\hline
\multirow{2}{*}{\textbf{Design Approach}} & \multicolumn{3}{c|}{\begin{tabular}[c]{@{}c@{}}\textbf{Energy Consumption}\\ \textbf{(pJ)}\end{tabular}} & \multicolumn{3}{c|}{\begin{tabular}[c]{@{}c@{}}$\textbf{Area}\times \textbf{Delay}$ \\ (\textbf{m\textsuperscript{2}}$\times$ \textbf{s})\end{tabular}} \\ 
\cline{2-7}
 & \textbf{D=1K} & \textbf{D=2K} & \textbf{D=8K} & \textbf{D=1K} & \textbf{D=2K} & \textbf{D=8K} \\ 
\hline
\textbf{\texttt{uHD}} per HV & 0.79 & 1.58 & 6.32 & 40.60$\times$10\textsuperscript{-12} & 81.20$\times$10\textsuperscript{-12} & 324.80$\times$10\textsuperscript{-12} \\ 
\hline
\textbf{\texttt{uHD}} per image (MNIST) & 113.76 & 227.52 & 910.08 & 5.83$\times$10\textsuperscript{-9} & 11.67$\times$10\textsuperscript{-9} & 46.69$\times$10\textsuperscript{-9} \\ 
\hline
Baseline per HV & 171.42 & 415.41 & 4023.82 & 11.79$\times$10\textsuperscript{-9} & 25.55$\times$10\textsuperscript{-9} & 230.33$\times$10\textsuperscript{-9} \\ 
\hline
Baseline per image (MNIST) & 24.68$\times$10\textsuperscript{3} & 59.80$\times$10\textsuperscript{3} & 57.94$\times$10\textsuperscript{4} & 1.70$\times$10\textsuperscript{-6} & 3.70$\times$10\textsuperscript{-6} & 33.17$\times$10\textsuperscript{-6} \\
\hline
\end{tabular}
\label{table_3}
\end{table*}

\ignore{
\begin{table}
\centering
\vspace{-0.5em}
\caption{Overall Design Performance}
\setlength{\tabcolsep}{3.pt}
\vspace{-1em}
\begin{tblr}{
  cells = {c},
  cell{1}{1} = {r=2}{},
  cell{1}{2} = {c=3}{},
  cell{1}{5} = {c=3}{},
  vlines,
  hline{1,3-5} = {-}{},
  hline{2} = {2-7}{},
      stretch = 0
}
\textbf{Design} & \textbf{ Energy Consumption $ (nJ) $} &  &  & \textbf{ Area ($mm^2) $} &  & \\
 & $1K $ & $2K $ & $8K $ & $1K $ & $2K$ & $8K$\\
\textbf{\texttt{uHD}} & $0.148$ & $0.297$ & $1.188$ & $31$\textcolor[rgb]{0.753,0,0}{} & 100\textcolor[rgb]{0.753,0,0}{} & $156$\\
 Baseline HDC & $24.75$ & $59.93$ & $579.9$ & $133$ & $269$ & $1079$
\end{tblr}
\label{table_4}
\vspace{-1em}
\end{table}
}

\section{Design Evaluation and Results}
\label{design_evaluation}

We evaluate the performance, hardware cost, energy consumption, and area$\times$delay of \textbf{\texttt{uHD}} compared to the baseline architecture and the prior state-of-the-art (SOTA).  
We first utilize the standard MNIST dataset for accuracy evaluations~\cite{726791}. 
We specifically compare the hardware costs 
of the baseline and \textbf{\texttt{uHD}} architecture 
for the hypervector generation process. The baseline design follows the dynamic and independent training target. 
Linear-feedback shift register (LFSR) modules are used for hypervector generation 
in the baseline design. Table~\ref{table_3} compares the energy consumption and area-delay product as important metrics to evaluate the hardware efficiency of the proposed design. 
We note that even though the iterative design is required for the baseline design (like $i$=100 different attempts to get the best-performing hypervectors), we credit it 
by assuming that hypervectors are the best, and only a single-run is 
sufficient for high accuracy. 
Thus, we 
provide a fair hardware comparison with respect to \textbf{\texttt{uHD}}. However, for a realistic 
baseline training 
on an edge device, more iterations are needed for high accuracy, which accordingly increases the energy consumption of the baseline design. 
We estimate the energy consumption for generating each $\boldsymbol{P} \times \boldsymbol{L}$ hypervector in the baseline HDC (Fig.~\ref{conventinal_steps}(b)) and for each $\boldsymbol{L}$ hypervector in the \textbf{\texttt{uHD}} design (Fig.~\ref{propose_sobol_index}). As can be seen in the reported numbers, our proposed \textbf{\texttt{uHD}} is more hardware-efficient than the baseline HDC.  


Table~\ref{survey_access} provides a comparative analysis of the SOTA 
HDC architectures 
implemented on a central processing unit or microprocessor. A thorough benchmarking is outlined in the HDC surveys 
by Hassan et al. \cite{9354795} and Chang et al. \cite{JETCASsurvey}. This table ranks the top energy-efficient frameworks for overall architecture (including \textit{hypervector generation}, \textit{binding}, \textit{bundling}, and \textit{binarization}) from the aforementioned surveys and contrasts them with our proposed architecture, which benefits from 
a novel approach for hypervector generation. 
As it can be seen, our proposed HDC architecture provides remarkable energy efficiency by exploiting \UC. 

\begin{table}
\centering
\caption{Energy Efficiency over Baseline Architectures~\cite{9354795, JETCASsurvey}}
\renewcommand{\arraystretch}{1.2}
\setlength{\tabcolsep}{5.5pt}

\begin{tabular}{|c|c|c|} 
\hline
\textbf{HDC Framework} & \textbf{Platform} & \textbf{Energy Efficiency} \\ 
\hline
Semi-HD \cite{SemiHD} & Raspberry Pi & 12.60$\times$ \\ 
\hline
Voice-HD \cite{voiceHD} & Central Processing Unit & 11.90$\times$ \\ 
\hline
tiny-HD \cite{BehnamIoT} & Microprocessor & 11.20$\times$ \\ 
\hline
PULP-HD \cite{PULP-HD} & ARM Microprocessor & 9.9$\times$ \\ 
\hline
Hierarchical-MHD \cite{Imani_Hierarchical} & Central Processing Unit & 6.60$\times$ \\ 
\hline
AdaptHD \cite{adapthd} & Raspberry Pi & 6.30$\times$ \\ 
\hline
Laelaps \cite{sandwich} & Central Processing Unit & 1.40$\times$ \\ 
\hline
\textbf{This work} & \textbf{ARM Microprocessor} & \textbf{31.83}$\times$ \\
\hline
\end{tabular}
\justify{\scriptsize{
In a comprehensive survey conducted in~\cite{9354795} and~\cite{JETCASsurvey}, several HDC frameworks were benchmarked based on their energy efficiency in comparison to reference baseline models. This table lists the top energy-efficient architectures from \cite{9354795} and \cite{JETCASsurvey}. All frameworks, including ours, report the overall (including memory read/write, hypervector generation, binding, and bundling) system energy consumption.
}}
\label{survey_access}
\end{table}

\begin{table}
\centering
\caption{MNIST Accuracy Performance ($\%$) of \textit{Baseline HDC} and \textbf{\texttt{uHD}}}
\setlength{\tabcolsep}{5.3pt}
\renewcommand{\arraystretch}{1.2}

\begin{tabular}{|c|c|c|c|c|c|c|c|} 
\hline
\multirow{2}{*}{\textbf{\textit{D }}} & \multicolumn{6}{c|}{{\textbf{Baseline HDC (Average)}}} & \textbf{\texttt{uHD}}  \\ 
\cline{2-8}
 & \textbf{\textit{i=1}} & \textbf{\textit{i=1..5}} & \textbf{\textit{i=1..20}} & \textbf{\textit{i=1..50}} & \textbf{\textit{i=1..75}} & \textbf{\textit{i=1..100}} & \textbf{\textbf{\textit{i=1}}} \\ 
\hline
\textbf{\textit{1K}} & 82.93 & 83.60 & 83.49 & 82.70 & 82.88 & 82.63 & \textbf{84.44} \\ 
\hline
\textbf{\textit{2K}} & 86.24 & 86.58 & 87.05 & 86.35 & 86.37 & 86.53 & \textbf{87.04} \\ 
\hline
\textbf{\textit{8K}} & 88.30 & 88.55 & 88.25 & 88.13 & 88.14 & 88.13 & \textbf{88.41} \\
\hline
\end{tabular}
\label{table_2}
\end{table}

\begin{figure}[t]
  \centering
  \includegraphics[width=\linewidth]{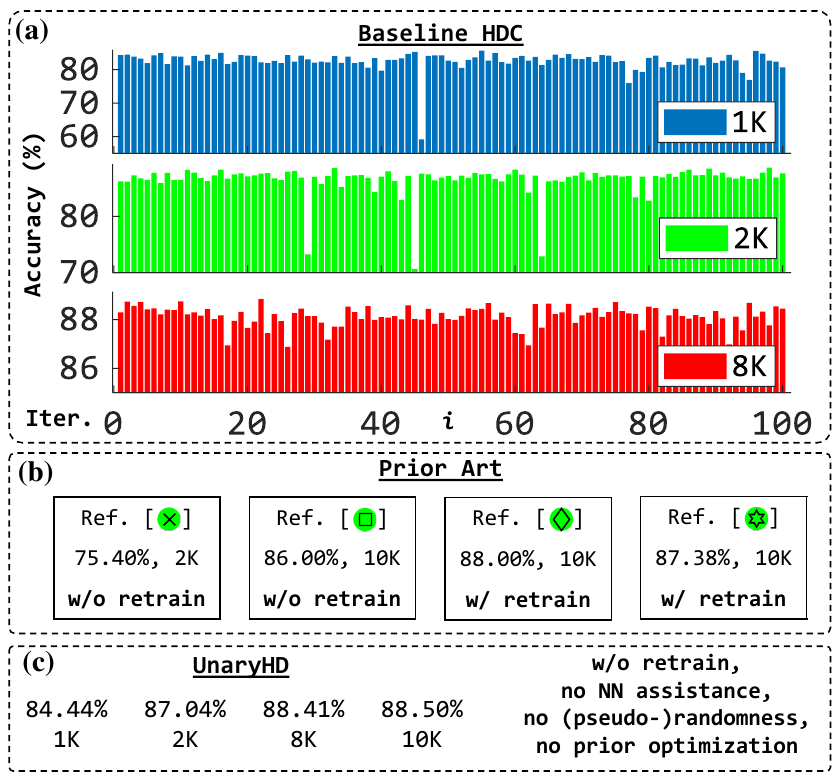}
  \vspace{-1.5em}
  \caption{Different accuracy monitoring of HDC designs. Accuracy fluctuations at each iteration of the baseline design with (pseudo-)randomness (a), MNIST dataset accuracy from prior HDC works; Ref.s \includegraphics[width=8pt,keepaspectratio=true]{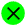} $\rightarrow$ \cite{8801933}, \includegraphics[width=8pt,keepaspectratio=true]{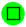} $\rightarrow$ \cite{9354795}, \linebreak \includegraphics[width=8pt,keepaspectratio=true]{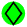} $\rightarrow$ \cite{9458526}, \includegraphics[width=8pt,keepaspectratio=true]{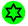} $\rightarrow$ \cite{QuantHD, duan2022braininspired}) (b), and \textbf{\texttt{uHD}} (c).}
  \label{random_plot}
  \label{fluctuate_random_acc}
\end{figure}

Table~\ref{table_2} compares the accuracy performance of the baseline HDC and \textbf{\texttt{uHD}}. The baseline architecture is monitored at different iterations of generating hypervectors ($\boldsymbol{P}$ and $\boldsymbol{L}$). At each random hypervector assignment in the training phase ($i$=$1...100$), the test accuracy is recorded. The table reports the average accuracy values at different checkpoints of $i$. 
\textbf{\texttt{uHD}} utilizes LD Sobol sequences and completes its deterministic hypervector (only $\boldsymbol{L}$) assignment in a single iteration ($i$=$1$). The MNIST dataset is segmented to separate the 
training and testing images, and for the sake of fair accuracy comparison between the two designs, there is no retaining, no neural network (NN) assistance, and no prior optimizations. 
Some prior 
work 
heavily 
relies on these optimizations; however, these 
optimizations and the use of other machine learning (ML) techniques 
affect the cost 
of the training hardware. The impact of using 
random vectors in the baseline HDC is reported in Fig.~\ref{fluctuate_random_acc}(a). The fluctuations 
in the testing accuracy underscore the importance of having an iterative process for selecting the best vectors. 
Fig.~\ref{fluctuate_random_acc}(b) reports the accuracy of prior SOTA HDC systems  
(the ones without 
NN assistance, complex optimizations in training, or multi-models - only with (w/) or without (w/o) the retraining efforts) in performing MNIST classification. As can be seen, 
\textbf{\texttt{uHD}} with 
single-pass learning 
achieves better accuracy compared to the baseline and SOTA 
designs. 

To extend our evaluations, we utilized various image-based datasets, including CIFAR-10, BloodMNIST, BreastMNIST, FashionMNIST, and SVHN~\cite{huggingFace}. Table~\ref{table_datasets} presents the accuracy comparison between the baseline HDC and our proposed \textbf{\texttt{uHD}}. 
We note that these accuracy results were obtained without employing any optimization (e.g., 
retraining, NN assistance, or transfer learning). The findings demonstrate the effectiveness and versatility of \textbf{\texttt{uHD}} across different 
datasets, showcasing its potential for various machine vision 
applications.
The ISO-accuracy values presented above demonstrate that \textbf{\texttt{uHD}} exhibits superior hardware efficiency compared to conventional learning frameworks (ML, DNN), which often require resource-intensive hardware setups. As a result, \textbf{\texttt{uHD}} offers a more efficient and cost-effective solution to achieve the same accuracy level. 

\begin{table}
\centering
\caption{Accuracy (\%) Comparison of Baseline HDC and the \textbf{\texttt{uHD}} for different image datasets.}
\setlength{\tabcolsep}{3.7pt}
\renewcommand{\arraystretch}{1.2}

\begin{tabular}{|c|c|c|c|c|c|c|} 
\hline
\multirow{2}{*}{\textbf{Datasets}} & \multicolumn{2}{c|}{\textbf{D=1K}} & \multicolumn{2}{c|}{\textbf{D=2K}} & \multicolumn{2}{c|}{\textbf{D=8K}} \\ 
\cline{2-7}
 & \textbf{\textit{Ours}} & \textbf{\textit{Baseline}} & \textbf{\textit{Ours}} & \textbf{\textit{Baseline}} & \textbf{\textit{Ours}} & \textbf{\textit{Baseline}} \\ 
\hline
\textbf{CIFAR-10} & \textbf{39.29} & 38.21 & \textbf{40.28} & 40.26 & \textbf{41.97} & 41.71 \\ 
\hline
\textbf{Blood MNIST} & \textbf{53.05} & 48.52 & \textbf{55.86} & 51.20 & \textbf{57.88} & 51.82 \\ 
\hline
\textbf{Breast MNIST} & \textbf{68.59} & 68.47 & \textbf{69.23} & 69.11 & \textbf{71.15} & 70.93 \\ 
\hline
\textbf{Fashion MNIST} & \textbf{68.60} & 54.19 & \textbf{70.06} & 69.97 & \textbf{71.37} & 70.87 \\ 
\hline
\textbf{SVHN} & \textbf{60.29} & 60.06 & \textbf{61.73} & 61.24 & \textbf{62.87} & 62.82 \\
\hline
\end{tabular}
\label{table_datasets}
\vspace{-0.5em}
\justify{\scriptsize{
For the Baseline HDC, the $P$ and $L$ hypervectors were generated using conventional random sequence generation.


}}
\vspace{-0.5em}
\end{table}

\section{Conclusions}
\label{conclusions}
This study proposed a hybrid 
HDC system, \textbf{\texttt{uHD}}, by employing UBC in HDC for the first time. 
The new design 
simplifies hardware implementation, providing significant hardware cost savings compared to the baseline HDC. 
\textbf{\texttt{uHD}} utilizes 
LD sequences for deterministic and high-quality generation of hypervectors. 
It achieves higher accuracy compared to the baseline HDC while offering single-iteration training. We propose a novel hypervector generator by representing data in the unary domain and comparing data using a novel unary comparator. The impact of adding the proposed modules is studied by comparing the energy consumption of the proposed design with the baseline HDC at different design checkpoints.  
We hope that this work opens new avenues for HDC by employing the complementary advantages of 
emerging computing technologies such as SC and UBC. 

\section*{Acknowledgments}
This work was supported in part by National Science Foundation (NSF) grant \#2019511, the Louisiana Board of Regents Support Fund \#LEQSF(2020-23)-RD-A-26, and generous gifts from Cisco, Xilinx, and Nvidia.

\bibliographystyle{IEEEtran}
\bibliography{bibliography,Hassan}


{
\begin{IEEEbiography}[{\includegraphics[width=1in,height=1.25in,clip,keepaspectratio]{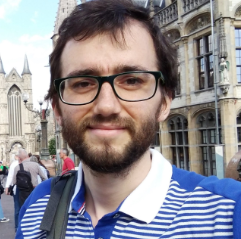}}]{Sercan Aygun} (S’09-M’22) received a B.Sc. degree in Electrical \& Electronics Engineering and a double major in Computer Engineering from Eskisehir Osmangazi University, Turkey, in 2013. He completed his M.Sc. degree in Electronics Engineering from Istanbul Technical University in 2015 and a second M.Sc. degree in Computer Engineering from Anadolu University in 2016. Dr. Aygun received his Ph.D. in Electronics Engineering from Istanbul Technical University in 2022. Dr. Aygun’s Ph.D. work has appeared in several Ph.D. Forums of top-tier conferences, such as DAC, DATE, and ESWEEK. He received the Best Scientific Research Award of the ACM SIGBED Student Research Competition (SRC) ESWEEK 2022 and the Best Paper Award at GLSVLSI'23. Dr. Aygun's Ph.D. work was recognized with the Best Scientific Application Ph.D. Award by the Turkish Electronic Manufacturers Association. He is currently a postdoctoral researcher at the University of Louisiana at Lafayette, USA. He works on emerging computing technologies, including stochastic and hyperdimensional computing in computer vision and machine learning. \end{IEEEbiography}


\begin{IEEEbiography}[{\includegraphics[width=1in,height=1.25in,clip,keepaspectratio]{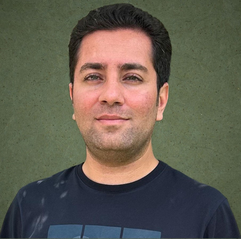}}]{Mehran Shoushtari Moghadam} (S’22) received the B.Sc. degree in Computer Engineering - Hardware and the M.Sc. degree in Computer Engineering - Computer Architecture from the University of Isfahan, Iran, in 2010 and 2016 respectively. He distinguished as one of the top-ranking students during both his B.Sc. and M.Sc. studies. He has more than 10 years of experience as a computer hardware and network specialist in the industry. He is currently a Ph.D. student at the School of Computing and Informatics, Center for Advanced Computer Studies, University of Louisiana at Lafayette, Lafayette, LA, USA. His research interests involve emerging and unconventional computing paradigms, including energy-efficient stochastic computing, real-time and highly-accurate brain-inspired computing, and hardware security. \end{IEEEbiography}


\begin{IEEEbiography}[{\includegraphics[width=1in,height=1.25in,clip,keepaspectratio]{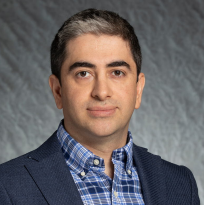}}]{M. Hassan Najafi} (S’15-M’18-SM'23) received the B.Sc. degree in Computer Engineering from the University of Isfahan, Iran, the M.Sc. degree in Computer Architecture from the University of Tehran, Iran, and the Ph.D. degree in Electrical Engineering from the University of Minnesota, Twin Cities, USA, in 2011, 2014, and 2018, respectively. He is currently an Assistant Professor with the School of Computing and Informatics, University of Louisiana, LA, USA. His research interests include stochastic and approximate computing, unary processing, in-memory computing, and hyperdimensional computing. He has authored/co-authored more than 75 peer-reviewed papers and has been granted 5 U.S. patents with more pending. In recognition of his research, he received the 2018 EDAA Outstanding Dissertation Award, the Doctoral Dissertation Fellowship from the University of Minnesota, and the Best Paper Award at the ICCD’17 and GLSVLSI'23. Dr. Najafi has been an editor for the IEEE Journal on Emerging and Selected Topics in Circuits and Systems. \end{IEEEbiography}

}

\balance

\end{document}